\documentclass[aps,prb,twocolumn,superscriptaddress,showpacs]{revtex4-1}
\usepackage{graphicx}
\usepackage{dcolumn}
\usepackage{bm}
\usepackage{color}
\usepackage{amsmath,amssymb}

\newcommand{\ket}[1]{\left| #1 \right>} 
\newcommand{\bra}[1]{\left< #1 \right|} 
\newcommand{\tip}{ {\rm tip}}
\newcommand{\imp}{ {\rm imp}}
\begin{document}
\title{From Kondo to local singlet state in graphene nanoribbons with magnetic impurities}

\author{G. S. Diniz}
\affiliation{Curso de F\'isica, Universidade Federal de Goi\'as,
Jata\'i, GO 75801-615, Brazil}
\author{G. I. Luiz}
\affiliation{Instituto de F\'isica,  Universidade Federal Fluminense,
Niter\'oi, RJ 24210-340, Brazil}
\affiliation{Instituto de F\'isica,  Universidade Federal  de
Uberl\^andia,Uberl\^andia, MG 38400-902, Brazil}
\author{A. Latg\'e}
\affiliation{Instituto de F\'isica,  Universidade Federal Fluminense,
Niter\'oi, RJ 24210-340, Brazil}
\author{E. Vernek}
\affiliation{Instituto de F\'isica,  Universidade Federal  de
Uberl\^andia,Uberl\^andia, MG 38400-902, Brazil}

\date{\today}

\begin{abstract}

A detailed analysis of the Kondo effect of a magnetic impurity in  a
zigzag graphene nanoribbon is addressed. An adatom is coupled to the graphene
nanoribbon  via a hybridization amplitude $\Gamma_\imp$ in a \emph{hollow}
or \emph{top} site configuration. In addition, the adatom is also weakly
coupled to a metallic STM tip by a hybridization function $\Gamma_\tip$ that
provides a Kondo screening of its magnetic moment. The entire system is
described by an Anderson-like Hamiltonian whose low-temperature
physics is accessed by employing  the numerical renormalization group
approach, which allows us to obtain the thermodynamic properties used to compute
 the Kondo temperature of the system. We find two screening regimes when the
adatom is close to the edge of the
zigazag graphene nanoribbon: (1) a weak coupling regime ($\Gamma_\imp\ll
\Gamma_\tip$), in which the edge states produce an enhancement of the Kondo
temperature $T_K$  and (2) a strong coupling regime ($\Gamma_\imp\gg \Gamma_
\tip$), in which a local singlet is formed, in detriment to the Kondo screening
by the STM tip. These two regimes can be clearly distinguished  by the
dependence of their characteristic temperature $T^*$  on the coupling
between the adatom and the carbon sites of the graphene nanoribon ($V_\imp$). We
observe that in the weak coupling regime $T^*$  increases exponentially with
$V_\imp^2$. Differently,  in the strong coupling regime, $T^*$ increases
linearly with  $V_\imp^2$.

\end{abstract}

\maketitle

\section{Introduction}
\begin{figure}[th!]
\begin{center}
\includegraphics[scale=0.74]{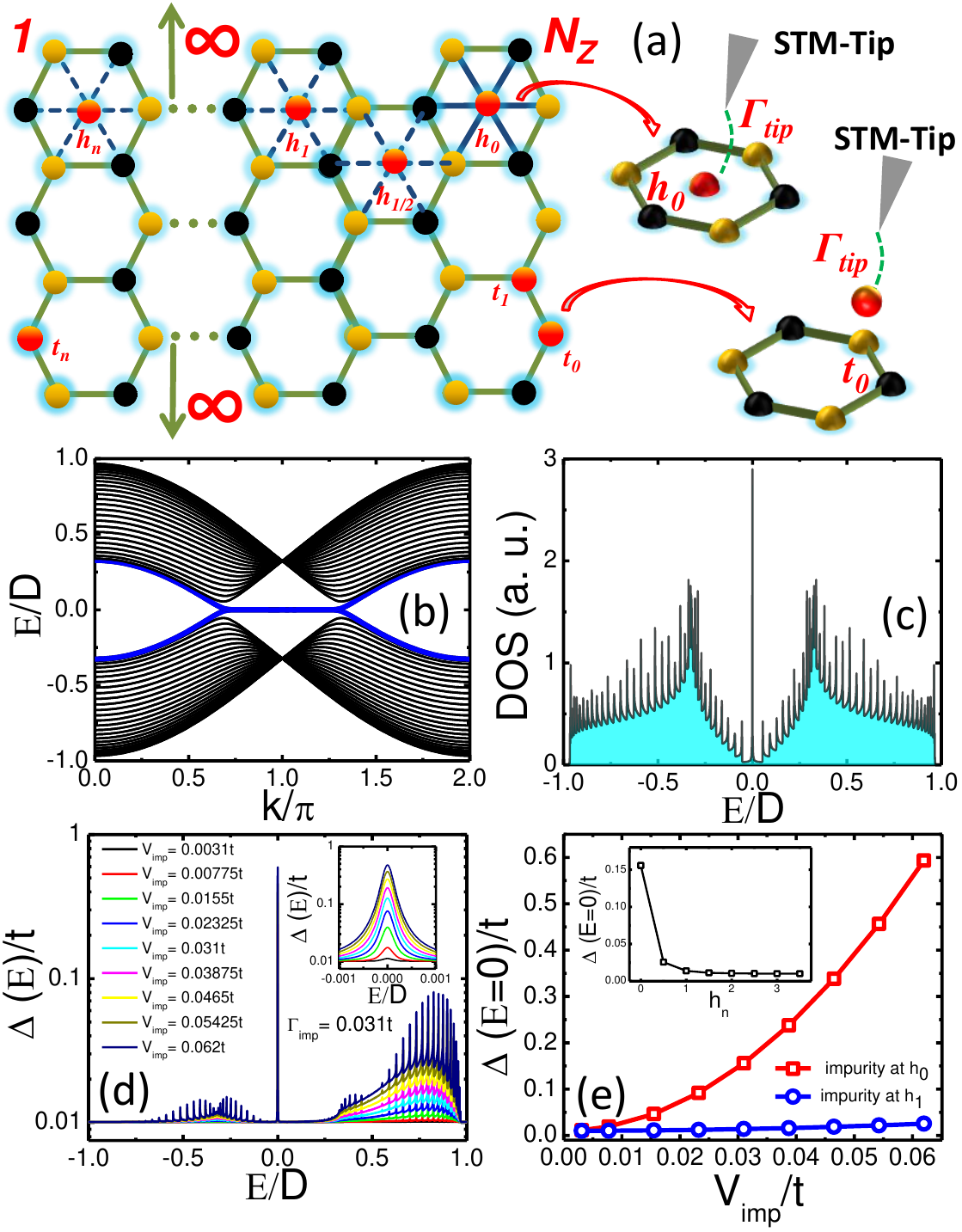}
\caption{(a) Schematic representation of $N_{z}$-ZGNR
with an impurity located at the hollow site position $h_{n}$ or at the top
site position $t_{n}$. The drawing on the right
represents a STM tip on the top of the impurity adatom with a coupling
strength $\Gamma_{\tip}$. (b) Electronic structure of a pristine 26-ZGNR
highlighting the edge states (blue line). (c) DOS as function of $E$ for
the pure graphene. (d) Hybridization function $\Delta(E)$ for different
values of impurity coupling strength $V_{\imp}$ at $h_{0}$ position. Inset: zoom close to
$E=0$ that shows the evolution of the $\Delta(E=0)$ peak as $V_{\imp}$
is enhanced. (e) The $\Delta(E=0)$ for two different impurity locations
$h_{0}$ and $h_{1}$ as function of $V_{\imp}$. Inset: $\Delta(E=0)$ vs different
adatom position $h_{n}$, for $V_{imp}=0.031t$.}
\label{fig1}
\end{center}
\end{figure}

Magnetic impurities embedded in a metallic environment exhibit the paradigmatic
many-body phenomena, the so-called Kondo effect (KE).\cite{Hewson-Kondo} Since
its  explanation in a seminal work by J. Kondo,\cite{Jkondo} this effect
has been studied in a variety of different physical systems in all dimensionalities
(tree-, two- and one-dimensional systems). \cite{Ernst,Jarillo,Nygard,Tarucha,Goldhaber}
The capability of manipulating atoms and molecules on metallic surfaces
with the aid of a STM tip has renewed the interest on the KE on reduced
dimensions, as we have witnessed  in the last twenty
years.\cite{2058-7058-14-1-28} In fact, the advent of the STM has paved the way
to a variety of  possibilities for investigating the KE in a controllable way
in many different
systems.\cite{Manoharan,PhysRevB.66.024431,PhysRevLett.97.266603} The great
number of theoretical and experimental studies has proved that the observable
physical signature of the KE depends drastically on the bare local density of
of states of the host system.

To understand the dependence of the Kondo physics with the density of states
of the free electrons surrounding the magnetic impurities, recall that the
physical mechanism underlying the KE is the \emph{dynamical screening} of the localized
magnetic moments of the impurities by the conduction electrons of the host
material. This screening is led by  an effective anti-ferromagnetic exchange
coupling $J$ between the impurity and the surrounding electrons that allows for
spin-flip scattering processes involving energies below $k_B T_K$, where $k_B$
is the Boltzmann constant and $T_K$ is the characteristic Kondo temperature.
Since these are spin-flip scattering events occurring at low temperatures, they
depend strongly on the low-energy density of states of the electrons nearby the
localized magnetic moments. This is why metallic systems with nearly constant
density of states  around the Fermi level  $E_F$
[$\rho(E)=\rho_F$] exhibits a typical Kondo temperature $T_K
\propto \exp(-1/\rho_F J)$, but can deviate drastically
from this expression if $\rho$ presents important features for $E $ close
to $E_F$. Within the single impurity Anderson model for  spin-$1/2$
magnetic impurity problems,\cite{PhysRev.124.41} the important quantity entering the expression
for $T_K$ is the ratio $U/\Delta(E_F)$, where $\Delta(E)\propto
\rho(E)$ is the effective hybridization function and $U$ is the
Coulomb repulsion energy at the impurity site. Again, we see that
$\rho(E)$ can greatly affect $T_K$. This is crucially important to
explain why it is observed enhanced $T_K$ in peaked
or in vanishing $T_K$ for pseudo-gaped effective hybridization function,
as discussed for a double quantum dot structure.\cite{PhysRevLett.97.096603}

A natural two-dimensional system exhibiting an interesting density
of state near the Fermi-level is graphene.
\cite{Novoselov22102004,Novoselov,RevModPhys.81.109,riseof} The Dirac cones of
the band structure lead to zero-gap density of states $\rho(E)\propto
|E|$ resulting in a rich phase diagram with interesting quantum phase
transitions.\cite{0034-4885-76-3-032501,PhysRevB.95.115408} There are
indeed great efforts devoted to the Kondo physics in graphene with an
impurity coupled \cite{0034-4885-76-3-032501,1367-2630-15-5-053018,
PhysRevB.77.045417,PhysRevB.81.115427,PhysRevB.83.165449,PhysRevB.84.165105,
PhysRevB.88.155412,PhysRevB.88.201103,PhysRevB.95.115408,PhysRevB.90.035426} or a
vacancy (defect) in the graphene lattice \cite{Fuhrer,PhysRevB.83.241408,JPSJ.81.063709,
PhysRevB.88.075104,PhysRevB.90.201101} in the recent years.
Surprisingly, much less attention has been paid to the KE on graphene
nanoribbons.\cite{PhysRevB.89.035424} A graphene nanoribbon (GNR) is
formed by breaking the translational symmetry of a graphene sheet in one
particular direction. There are two common directions for \emph{cutting} the
graphene with well defined edge shapes: zigzag (ZGNR) and armchair
(AGNR). ZGNRs are particularly  interesting because around $k=\pi$ they
exhibit simultaneously dispersive bulk  and bound edge
states.\cite{Fujita,1367-2630-11-9-095016} These bound edge states render a
strongly peaked local density of states, as depicted in Fig.~\ref{fig1}(c).

Because of the sharp peak in the local density of states, if we place
a magnetic impurity near one edge of the ZGNR one can expect an
important modification on the Kondo physics of the system. In particular,
in view of the discussion above, one can expect this peak to
enhance the Kondo temperature of the system.  Recent theoretical DFT
calculations predict that adatoms are possible generators of
localized  magnetic moments in graphene and GNR either at hollow  and top site
positions.
\cite{PhysRevB.77.195434,PhysRevB.79.245416,1367-2630-12-5-053012,PhysRevB.84.245411,PhysRevLett.102.126807,
PhysRevB.84.195444,PhysRevB.89.035424,Yuliang}
Motivated by these findings, in this work we are interested in
investigating KE of a magnetic impurity placed at two distinct
positions in a ZGNR: hollow site and top
site.\cite{PhysRevLett.110.136804,PhysRevLett.111.236801,Gonzalez-Herrero437}
We employ a numerical renormalization group  approach
(NRG),\cite{RevModPhys.47.773,RevModPhys.80.395} to address this
problem.  More precisely, using the NRG approach we calculate the entropy,
magnetic susceptibility (from which we can extract  the Kondo temperature with
the aid of the Wilson's criteria) of the system for an impurity  placed at the
hollow site for distinct locations along the transversal direction. We find a
strong enhancement of the $T_K$ when the impurity approaches the edge of the ZGNR. More
interestingly, in the strong coupling regime between the impurity and
the nearby carbon atoms, our calculations suggest that the impurity
magnetic moment forms a local singlet state with the edge state of the
ZGNR.

This paper is structured as follows: in Sec. II we
present the theoretical model describing the hybridization
function and the numerical renormalization group approach. In Sec. III
the numerical results  are discussed. Finally, in Sec. IV we
present our conclusions.

\section{Theoretical Model and method}

The system is schematically illustrated in Fig. \ref{fig1} (a), where a $N_{Z}$-ZGNR, with
the index $N_{Z}$ standing to the number of zigzag chains in the transversal
direction. To model the single magnetic impurity hosted in the GNR (at the
hollow site or at the top site position), we
have used the Anderson-like Hamiltonian, \cite{PhysRev.124.41}
\begin{align}
\label{H1}
H \!=\! H_{\rm GNR} + H_{\rm imp}+H_{\rm tip} + H_{\rm GNR-imp} +H_{\rm
imp-tip},
\end{align}
where the first term describes the GNR that is modeled by a tight-binding
Hamiltonian
\begin{eqnarray}
H_{\rm
GNR}=\sum_{i\sigma}(\varepsilon_0 - \mu)c^\dag_{i\sigma}c_{i\sigma}-t\sum_{\langle
i,j\rangle,\sigma}c_{i\sigma}^{\dagger} c_{j\sigma}
\end{eqnarray}
in which the operator $c_{i\sigma}^{\dagger}$ ($c_{i\sigma}$) creates
(annihilates) an electron with energy $\varepsilon_0$ and  spin $\sigma$ in the
$i$-th carbon site of the GNR, and $\mu$ is the chemical potential that can be
externally tuned by a back gate. The matrix element $t$ allows
the electron to hop between nearest neighbor carbon sites. \cite{RevModPhys.81.109} The second term in
Eq.~\eqref{H1} describes the single level Anderson impurity that is modeled by
the interacting Hamiltonian $H_{\rm imp} = \varepsilon_{d}n_{d\sigma}  +
Un_{d\uparrow}n_{d\downarrow}$, where $d^\dagger_{\sigma}$ ($d_{\sigma}$)
creates (annihilates) an electron with energy $\varepsilon_d$ and spin $\sigma$ at
the impurity site, $U$ is the on-site Coulomb interaction and
$n_d=n_{d\uparrow}+n_{d\downarrow}$ (with $n_{d\sigma}\equiv
d^\dagger_{\sigma}d_{\sigma}$) is the total number operator for the impurity
electrons.  The third term in Eq.~\eqref{H1} describes the STM tip modeled
by the Hamiltonian $H_{\rm tip}=\sum_{\bf k}\varepsilon_{\bf k}c^\dagger_{\bf
k\sigma}c_{\bf k\sigma}$, where $c^\dagger_{\bf k\sigma}$ ($c_{\bf k\sigma}$)
creates (annihilates) an electron with momentum ${\bf k}$ and spin $\sigma$ in the
STM tip. Finally, the last two terms of the Eq.~\eqref{H1} couples the impurity
to the GNR and to the STM tip, respectively. They are, respectively, given by
\begin{align}
\label{H2}
H_{\rm GNR-imp} =
\sum_{j,\sigma}V_{j}c_{j\sigma}^{\dagger}d_{\sigma},
\end{align}
and
\begin{eqnarray}
\label{H3}
H_{\rm imp-tip}=\sum_{{\bf k}\sigma}\left(V_{\bf k}c^\dagger_{\bf k
\sigma}d_{\sigma}+{\rm H.c.} \right).
\end{eqnarray}
In Eq.~\eqref{H2}  $V_{j}$ represents the impurity coupling
amplitude to the neighboring carbon atoms (later, we set $V_{j}\equiv
V_{\imp}$). For the impurity located at the
hollow site position $h_{n}$, the sum in $j$ runs over the six carbon atoms closest to the
impurity, while for the top site position $t_{n}$, it only corresponds to the single carbon atom
which the impurity is coupled to.

\subsection{Hybridization function}
The implementation of the NRG to determine the Kondo
temperature of the system requires first the determination of the
hybridization function, $\Delta(E)$, of the impurity. We do it by using
the Green's function method to the non-interacting case ($U=0$). As the
impurity is coupled to both the GNR and to the STM tip, we write
$\Delta(E)=\Delta_{\rm
tip}(E)+\Delta_{\rm GNR}(E)$. To obtain $\Delta_{\rm tip}(E)$ we
model the STM tip by a constant density of state  $\rho_{tip}$ and assume  a
coupling, $V_{\bf k}=V_{\rm tip}$ (independent of ${\rm k}$), so that we can write
$\Delta_{\rm tip}(E)=\pi V_{\rm tip}^2\rho_{\rm tip}\equiv \Gamma_{\rm
tip}$. To obtain $\Delta_{\rm GNR}(E)$ we have implemented the standard
surface Green's function approach.\cite{Nardelli,Sancho} The GNR is then
divided into three regions: left lead, central region (where the impurity adatom
is located) and right lead. The retarded Green's function matrix of the central
region is
${ G}_{C}(E)=\left(E + i\eta -{ H}_{C}-{ \Sigma}_{L}-{ \Sigma}_{R}
\right)^{-1}$ (omitting the spin indices), with $E$ being the energy of the
injected electron (the Fermi energy at a given doping) and $\eta\rightarrow 0$.
Here, $H_{C}$ represents the Hamiltonian describing the central region and
$\Sigma_{L/R}$ are the self-energies that describe the influence of  the
left/right leads. Explicitly, $\Sigma_{l}=H_{lC}^{\dagger}g_{l}H_{lC}$, where
$g_{l}$ is the Green's function for the $l=L,R$ semi-infinite lead obtained through an
iterative procedure of the tight-binding Hamiltonian,\cite{Nardelli} and
$H_{lC}$ couples each lead to
the central region. With the Green's function, we can obtain the self energy
of the impurity site $\Delta_{\rm GNR}(E)={\rm Im}[ G^{-1}_{C}(E)
]_{NN}$, where $N$ represents the impurity site inside the central region.

Differently from the graphene case, in which for the impurity located at the
top site $\Delta(E)$ is a linear function of
$\vert E\vert$ (or as $\vert E\vert^3$ for the hollow site),
\cite{PhysRevLett.106.016801,PhysRevB.95.115408}
in the ZGNR $\Delta(E)$ displays a much more complex dependence on $E$.
In the zigzag GNR, for instance, the presence of the edge state dramatically
alters the hybridization function $\Delta(E)$. To illustrate this, in
Fig.~\ref{fig1}(b) we show the electronic structure of a 26-ZGNR (where the
$26$ stands for the number of zigzag chains along the transverse direction). The
blue curve corresponds to the edge states. Note that it exhibits a flat
zero energy plateau around $k=\pi$. This plateau gives rise
to a sharp zero-energy peak in the DOS of the pristine 26-ZGNR (in the absence
of the impurity), $\rho_{\rm GNR}^{(0)}(E)={\rm Im}\, {\rm
Tr}[G_c^{(0)}(E)]$, as shown in Fig.~\ref{fig1}(c).

Since the sharp contribution to $\rho^{(0)}_{\rm GNR}(E)$ is located at the
edges of the ZGNR, for the impurity coupled to the carbon atoms close to the
edges a strong enhanced hybridization is observed at $E=0$, as shown in
Fig.~\ref{fig1}(d). Also, we can  notice that several satellite peaks appear,
as a consequence of the van Hove singularities in the DOS of the ZGNR.\cite{Fujita,1367-2630-11-9-095016}
Interestingly, note that while $\rho_{\rm GNR}^{(0)}(E)$
[Fig.~\ref{fig1}(c)] is particle-hole symmetric $\Delta(E)$
[Fig.~\ref{fig1}(d)] is not. This is because while the former is calculated in
the absence of the impurity, the last is defined with the impurity coupled to
the GNR, which for the hollow site breaks particle-hole symmetry of the
system, a known behavior for non-bipartite
lattices.\cite{Muller-Hartmann1995,PhysRevLett.92.216401}
By increasing the impurity coupling $V_{\rm imp}$, the zero-energy peak of
$\Delta(E=0)$ also increases. This is better appreciated in the inset of
Fig.~\ref{fig1}(d). In Fig.~\ref{fig1}(e) we show $\Delta(E)$ vs $V_{\rm
imp}$ for the  impurity placed at $h_{0}$ (red squares) and $h_1$ (blue
circles). We first note that $\Delta(E=0)$ is much
larger for the impurity at the position $h_0$ than in position $h_1$, see inset of
Fig.~\ref{fig1}(e), for $V_{imp}=0.031t$ \cite{adatom}, which is consistent with the expected decay of the edge state
wave function across the ribbon width. Moreover, we can observe that $\Delta(0)\propto
V_{\rm imp}^{2}$, similar to the case of an impurity coupled to a metallic
surface. \cite{PhysRev.124.41}

For completeness, we have also analysed a metallic armchair graphene
nanoribon (AGNR) with an adatom at the hollow site, with a similar width as the
26-ZGNR. For this purpose, we choose the 47-AGNR, which means 47 $N_{A}$ (dimers
line) along the transversal direction:\cite{Fujita} however we did not find any
significant change in the local DOS close to the Fermi level, and
consequently for the hybridization function around $E=0$ for different adatom
impurity position along the transversal direction. This is a consequence of the
almost flat density of states around $E=0$ and the absence of edge states for
the AGNR.\cite{Fujita} For metallic AGNR
close to the Fermi level, we expect the Kondo physics to mimic the case of a magnetic impurity
hosted in a normal metal, where a nearly constant density of states is expected.

\subsection{Numerical renormalization group approach}
To provide the Kondo physics description of the GNR with an impurity adatom, we
use the Wilson's numerical renormalization group (NRG)
approach.~\cite{RevModPhys.47.773,RevModPhys.80.395} For this purpose, we set
the Wilson's discretization parameter as $\Lambda$= 2.0, retaining 2000
many-body states after each iteration, and using the $z$-averaging in the interval
$0.2\le z\le 1.0$, in steps of 0.2. \cite{PhysRevB.79.085106}
The entropy is obtained within the canonical ensemble as $S(T)=\beta\langle
H\rangle + \ln Z$, where $Z$ correspond to the number of occupied states.
Similarly, the magnetic moment is given by $Z^{-1}\Sigma_{n}[\bra{\Psi_{n}}
S_{z}^{2}\ket{\Psi_{n}} - \bra{\Psi_{n}} S_{z}\ket{\Psi_{n}}]\times e^{-\beta
E_{n}}$. It is important to mention that we seek for the entropy
$S_\imp$, and magnetic moment $\chi_\imp$ which  correspond (approximately) to the
contribution of impurity to the entropy and to the magnetic moment, and are
defined as the difference of the thermodynamical quantities computed for
total Hamiltonian $H$ (with the impurity) and with $H_{0}$ (without the
impurity). Further technical details can be found, for instance, in
Ref.~\onlinecite{RevModPhys.80.395} and references therein.

\section{Numerical Results}
The following results are for the $N_{Z}$-ZGNR. To obtain our
numerical results, we choose the hopping $t$ such that the
half-bandwidth is unity, i.e. $D=1$, and can be used as our energy unity.
With this in mind, we use $U/t=1$ and the
$\Gamma_{\rm tip}=0.031t$ for all calculations. This choice, for $V_{\rm imp}=0$,
the hybridization of the impurity with the STM tip will render a very small
Kondo temperature (which depends essentially on the ratio $U/\Gamma_{\rm
tip}\approx 32.3$) that can be estimated by \cite{PhysRevLett.40.416}
$k_{B}T_K\approx \sqrt{\Gamma_{\rm tip} U} e^{-\pi U /8\Gamma_{\rm tip}}\approx
5.5\times10^{-7}t$ for $\delta=0$, where $\delta=\varepsilon_d+U/2$. Numerically, $T_K$
is obtained from the magnetic moment, following Wilson's criteria,\cite{RevModPhys.47.773}
$k_BT_K\chi(T_K)/(g\mu_{B}^{2})=0.0707$, where $g$ is the Lang\'e $g$-factor and $\mu_{B}$ is
the Bohr magneton. With this prescription we find (for $V_\imp=0$)
$k_BT_K\approx 9.6 \times10^{-8} t\equiv k_BT_K^{(0)}$ (this will be used later
to rescale our characteristic temperatures).

\begin{figure}[ht]
\begin{center}
\includegraphics[scale=0.5]{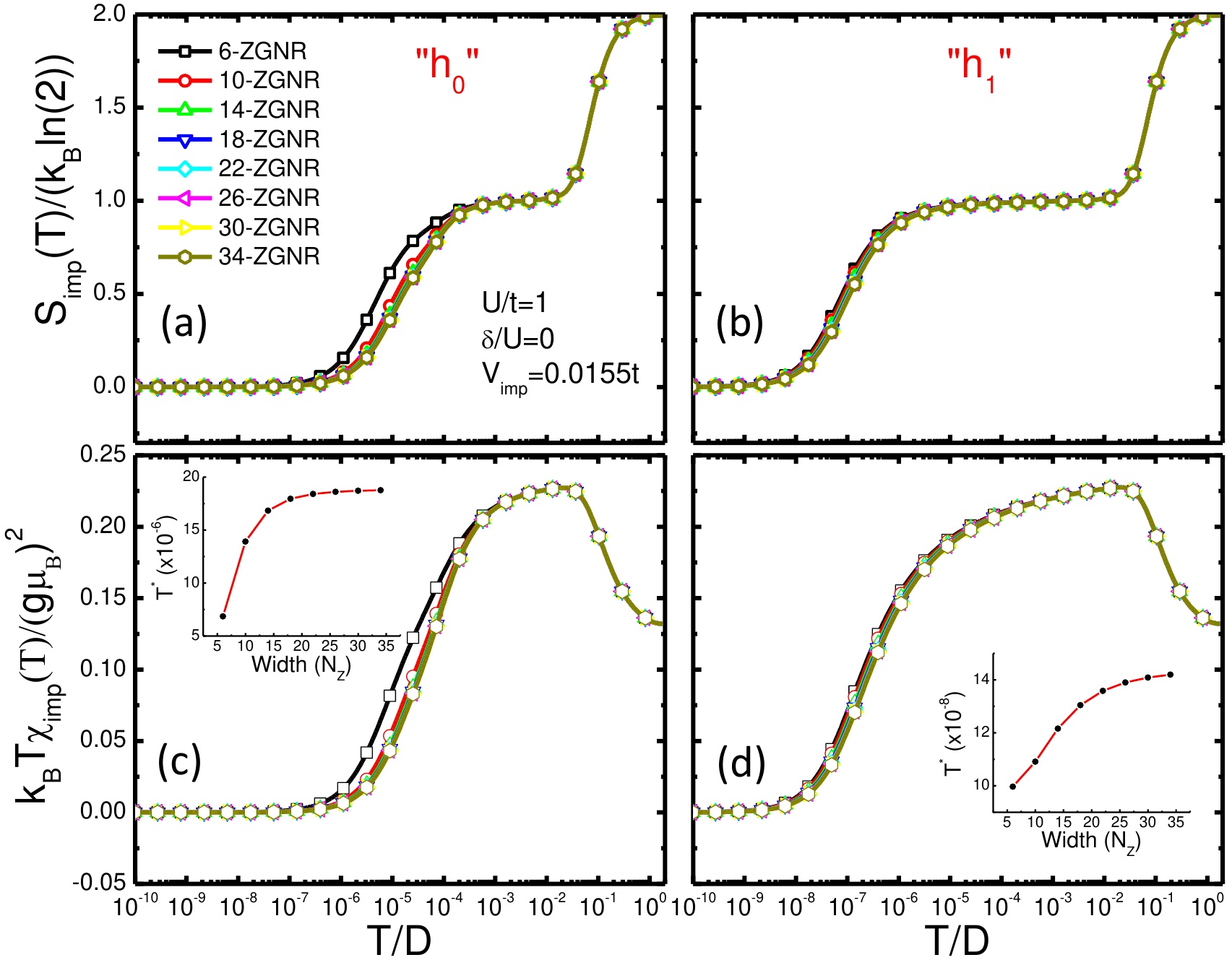}
\caption{Entropy (top) and magnetic moment (bottom) vs $T$ for the
impurity located at $h_{0}$ (left), $h_{1}$ (right), and for different widths of
the ZGNR. For all panels, $\delta/U=0$, $U/t=1$, and $V_{\imp}=0.0155t$.
The insets of panels (c) and (d) show the characteristic Kondo temperature
$T^{*}$ vs $N_Z$ for the corresponding adatom position.}
\label{fig2}
\end{center}
\end{figure}

\subsection{Hollow site adatom}
To study the effect of the edge state on the Kondo physics screening of the
system we first focus on the  hollow site position. In Fig.~\ref{fig2} we
show the entropy (top) and the magnetic moment (bottom) as function of
temperature for different $N_Z$-ZGNR for $V_{\imp}=0.0155t$, $\delta/U=0$, and
for the impurity located at $h_{0}$ (left) and $h_{1}$ (right). Overall, the
features observed for the $S_{\imp}(T)$ and $k_BT\chi_\imp(T)$ are similar to
those known for the traditional single Anderson impurity problem. As shown in
Fig.~\ref{fig2}(a) we see two drops of the $S_{\imp}(T)$ as $T$ decreases. The
first corresponds to the crossover from the free orbital to the local moment
regime  while the last one  corresponds to the quench of the local magnetic
moment by the conduction electrons which characterizes the onset of the KE.
Alike, the magnetic moment follows the same feature observed in the single
Anderson impurity embedded in a metal. What is interesting here is that, for a
given position of the impurity $h_0$ (left) or $h_1$ (right), as $N_Z$
increases, $S_{\rm imp}$ and $k_BT\chi_{\rm imp}$ drop to zero at higher
temperatures. This suggests that the \emph{sharpening} of the edge state by increasing
$N_Z$ enhances the Kondo temperature of the system. The insets of
Figs.~\ref{fig2}(c) and \ref{fig2}(d) show the characteristic temperature $T^{*}$ vs
$N_Z$ for the positions $h_0$ and $h_1$, respectively.
According to our previous discussion, note that for a given position of the
impurity, $h_0$ or $h_1$, $T^{*}$ with $N_Z$ increases and saturates
to a given value. Observe also that consistently with the exponential decay of
the wave function as it penetrates across the ZGNR width,\cite{Fujita} $T^{*}$
is much larger for the impurity placed at $h_0$ [\ref{fig2}(c)] than for the
situation in which it is placed at $h_1$ [\ref{fig2}(d)].  For the results
shown so far we fixed $V_{\imp}$ at small value ($0.0155t$). This assures that
even the strongest influence of the edge state of the ZGNR occurring for the
position $h_0$, it does not change the picture of the Kondo screening.

We address the question whether this picture remains in the regime in which the
magnetic impurity is more tightly connected to the ZGNR. To do so,
let us turn our attention to the dependence of magnetic moment suppression, as
we increase $V_{\imp}$ for a given impurity position. Here, we focus on the
$26$-ZGNR with the adatom at the hollow site position, for which $T^{*}$ is
almost converged [specially for the impurity adatom located at the edge, see
the insets of Figs.~\ref{fig2}(c) and \ref{fig2}(d)].  Since the drop of
$k_BT\chi_\imp$  corresponds  to a drop in $S_{\imp}$, it suffices to discuss
only one of them. Therefore, from now on we will discuss only $k_BT\chi_\imp$.
\begin{figure}[h]
\begin{center}
\includegraphics[scale=0.48]{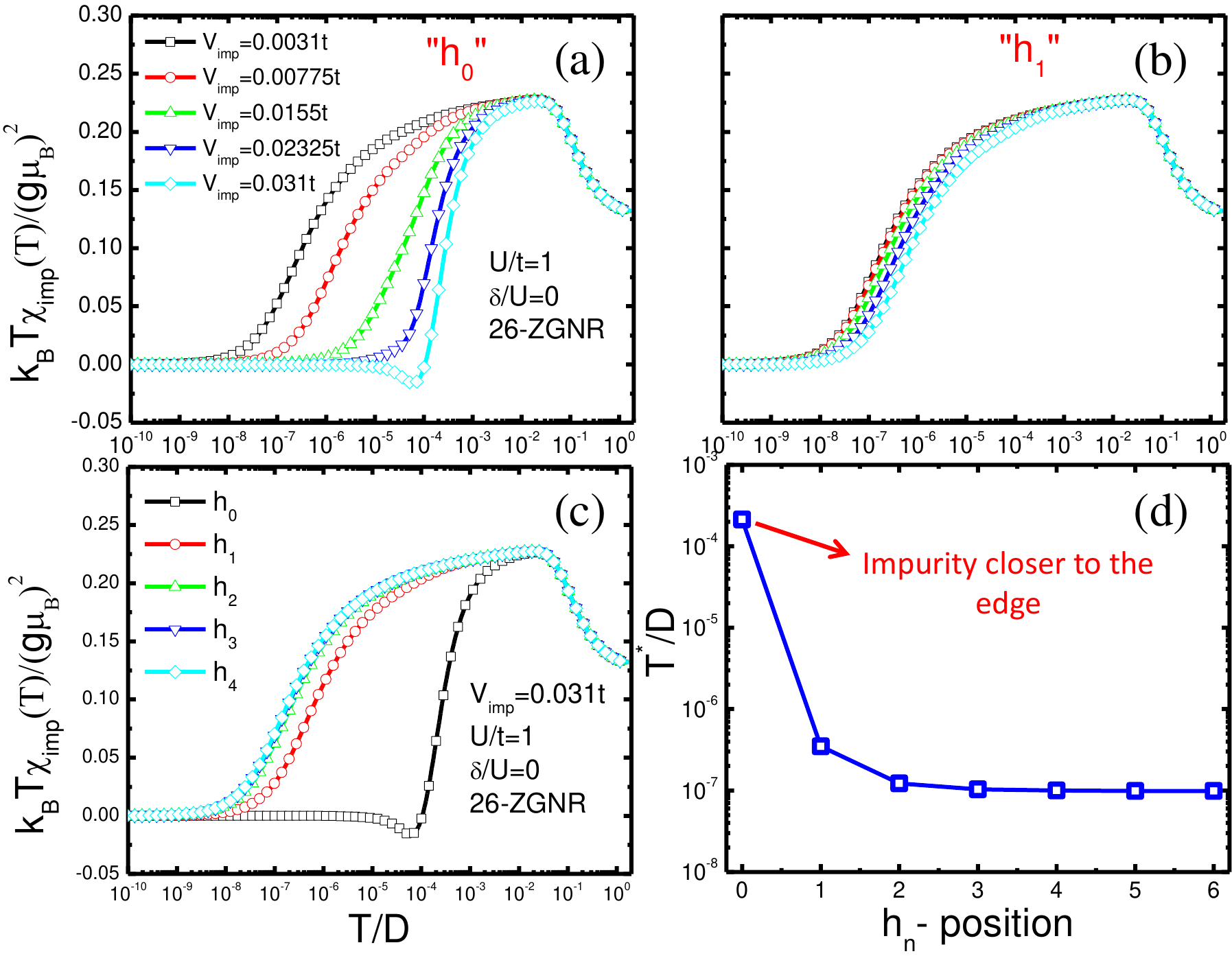}
\caption{Magnetic moment vs temperature for the
impurity located at $h_{0}$ (a)  and $h_{1}$ (b) and for various values of
$V_{\imp}$. (c) Magnetic moment vs temperature for different
positions of the  adatom $h_{n}$ with a fixed value of $V_{\imp}=0.031t$.
(d) Kondo temperature, $T^{*}$ vs position $h_{n}$ $(n=0,1,\ldots,6)$, with $V_{imp}=0.031t$.
[Notice that in this panel we have set the vertical axis in a log scale].}
\label{fig3}
\end{center}
\end{figure}
In Fig.~\ref{fig3}(a) and \ref{fig3}(b) we show the  impurity magnetic
susceptibility, $k_BT\chi_\imp$, vs $T$ for different values of $V_{\imp}$ for
the  impurity placed at $h_{0}$ and $h_{1}$, respectively.
First, by comparing the curves of Fig.~\ref{fig3}(a) with \ref{fig3}(b) we
observe  that the quenching of the magnetic moments is
much more strongly dependent of $V_\imp$ for $h_0$ than for $h_1$ position.
Again, this is because the closer is the impurity to the ZGNR edge, the
stronger is the hybridization of the impurity orbital with the edge state. This
also can be seen in Fig.~\ref{fig3}(c) showing $k_B\chi_\imp$ vs $T$ for different positions of the impurity on the $26$-GNR.
Note that the quench of the
magnetic moment occurs at much higher temperature for the position $h_0$ as
compared to the other positions. As the impurity is moved far away from the
edge, the curves of $k_BT\chi_\imp$ rapidly collapse on each other, approaching that
one for $V_\imp=0$. This is because far away from the edge the KE is
essentially due to the STM tip. Still using Wilson's
criteria, from the results of Fig.~\ref{fig3}(c) we extract the characteristic
temperature  $T^*$ below which the magnetic moment is quenched which is shown
in Fig.~\ref{fig3}(d) vs $h_{n}$, for fixed a $V_{\imp}=0.031t$. We can observe
that $T^*$ drops about three orders of magnitude as the impurity moves from
$h_0$ to $h_3$. A remarkable feature observed for large $V_\imp$ shown in
Fig.~\ref{fig3}(c), for instance, is that the shape of the magnetic moment for
$h_0$ (black curve) differs significantly from the other ones. The natural
question is now: is the suppression of the local magnetic moment for $h_0$
and large $V_\imp$ of the Kondo-like type?

\begin{figure}[ht]
\begin{center}
\includegraphics[scale=0.46]{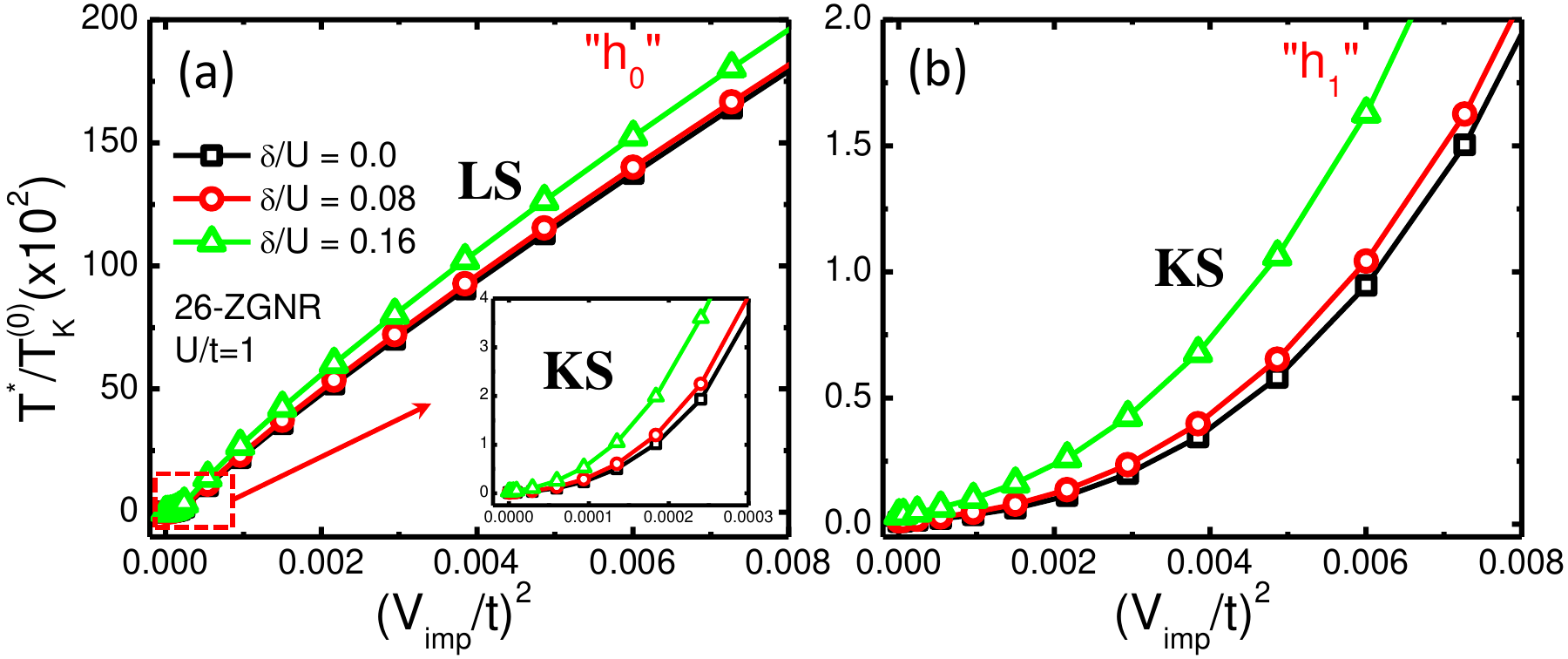}
\caption{$T^*/T^{(0)}_{K}$ as function of $(V_{\imp}/t)^2$ for (a) $h_{0}$ and
(b) $h_{1}$, for different $\delta/U$ ratios. The inset of panel (a) shows
a zoom of the region of small values of $(V_{\imp}/t)^2$. The
parameters used are in the legend of panel
(a) and $T^{(0)}_{K}=9.6\times10^{-8}t$.}
\label{fig4}
\end{center}
\end{figure}
To answer this question, we show in Fig.~\ref{fig4} the $T^*/T^{(0)}_{K}$ vs
$V^2_{\imp}$, for two different adatom location at the hollow site  position
$h_{0}$ in (a) and $h_{1}$ in (b), for different values of $\delta/U$. From
Fig.~\ref{fig4}(a) we can clearly distinguish two different regimes: for large
$V_\imp$ we note a linear behavior of $T^*$ with $V_\imp^2$ but for small
$V_\imp$ the dependence seems to be exponential. This is better seen in the
inset of Fig.~\ref{fig4}(a). To understand this, let us remember that when the
impurity is coupled to both the tip and the ZGNR, the effective hybridization is
given by $\Gamma=\Gamma_\tip +\Gamma_{\imp}$. Here, $\Gamma_{\imp}= a
V_\imp^ 2$, where $a$ is a constant that depends on the position of the
impurity, and the local density of states of the ZGNR. In the small $V_\imp$
regime we can use the expression for $T_K$ for $T^*$. Therefore,
\begin{eqnarray}
T^*\sim e^{-\pi U/8\Gamma}.
\end{eqnarray}
For $\Gamma_\imp \ll \Gamma_\tip$ we can write
\begin{eqnarray}
\frac{1}{\Gamma}\approx
\frac{1}{\Gamma_\tip}\left(1-\frac{\Gamma_\imp}{\Gamma_\tip} \right).
\end{eqnarray}
Thus we obtain
\begin{eqnarray}\label{T_smalV}
T^*\sim e^{-\pi U/8\Gamma_\tip}e^{(\pi
U/8\Gamma_\tip^2)\Gamma_\imp}\sim T_K^{(0)}e^{bV_\imp^2},
\end{eqnarray}
where $T_K^{(0)}$ is the Kondo temperature for $V_\imp=0$ and $b=\pi U
a/8\Gamma_\tip^ 2$. Note that the expression \eqref{T_smalV} is consistent with
the exponential behavior of $T^*$ shown in the inset of Fig.~\ref{fig4}(a).
This shows indeed that in the regime of small $V_\imp$ the quenching of the
magnetic moment is actually of the Kondo type, therefore we call it Kondo singlet (KS)
regime and so $T^*$ can be identified as $T_K$.

\begin{figure}[ht]
\begin{center}
\includegraphics[scale=0.45]{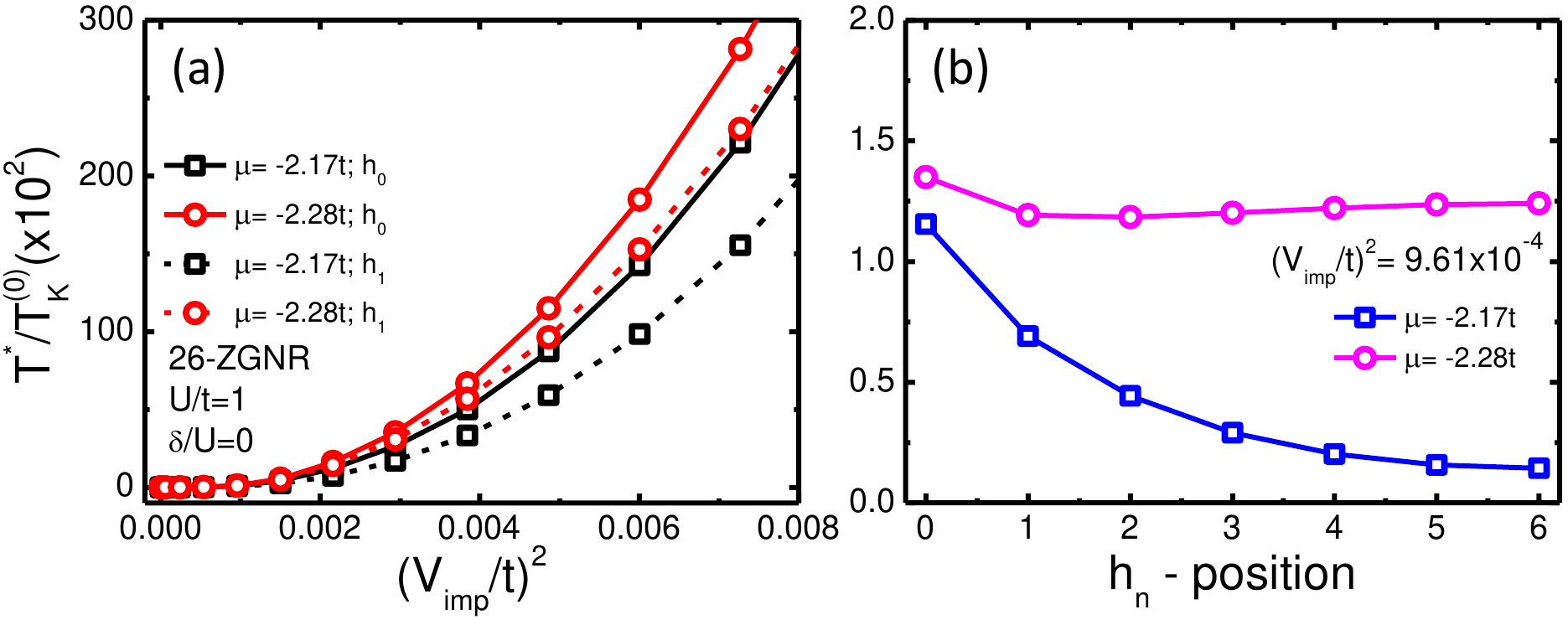}
\caption{(a) $T^*/T^{(0)}_{K}$ vs $(V_{\imp}/t)^2$ for  $\mu\!=\!-2,17t$ (black)
and $-2.28t$ (red) and $h_{0}$ (solid) and $h_{1}$ (dotted)
and. (b) $T^*$ vs adatom position $h_{n}$  $(n=0,\ldots,6)$, with
$(V_{\imp}/t)^2\!=\! 9.61\times10^{-4}$. The other parameters are
$\delta\!=\!0$,  and $U\!=\!t$. Here, $T^{(0)}_{K}\!=\!9.6\times10^{-8}t$.}
\label{fig5}
\end{center}
\end{figure}

In contrast, in the opposite regime, the behavior of $T^*$ can no
longer be understood within the picture described above. In this case, the
strong coupling between the  impurity and the bound edge state (for the
position $h_0$), induces the formation of local singlet (LS). Within this
picture, the impurity and the edge state can now be thought as a two hybridized
electronic levels with Coulomb repulsion $U$ in one of them. The energy gain to
form a LS state in this simple system is known to be $E_S= - 4V_\imp^2/U$. This
explain why in the strong impurity-ZGNR coupling regime $T^*\propto V_\imp^2$.
In this case, we prefer not to identify $T^*$ as $T_K$, since here the singlet
does not involve the Fermi sea as in the traditional KE. This is
actually akin to what was discussed by one of us in
Ref.~\onlinecite{PhysRevB.82.165304}.
If we now look at Fig.~\ref{fig4}(b) this linear behavior of $T^*$ with
$V_{\imp}^ 2$ is not observed (at least for the range of $V_\imp$ shown). This
is because at the position $h_1$ the influence of the edge state on the
impurity remains a small perturbation. Interestingly, as we observe the
similar behavior for the different curves of Fig.~\ref{fig5}, the two regimes
discussed above remain clearly distinguishable for $\delta \neq 0$.

As we have seen above, the behavior of $T^*\sim V_\imp^2$ could be nicely
understood in terms of the states bound to the edge of the ZGNR. One
can argue that if we place the chemical potential close to a van Hove
singularity this behavior would no longer be seen. This is because the van Hove
singularities are not directly associated to states bound to the edges. To
confirm this, we now consider the chemical potential $\mu$ at two different van
Hove singularities with a relative peak similar to the edge state, and calculate
$T^*$. In Fig.~\ref{fig5}(a) we show $T^*$ vs $V_\imp^2$ for
$\mu=-2.17 t$ (black) and $\mu=-2.28t$ (red) and for the two impurity
positions $h_0$ (solid lines) and $h_1$ (dotted lines). Notice that for
all cases $T^*$ increases exponentially with $V_\imp^2$, very similar to the
results shown in Fig.~\ref{fig4}(b) and in the inset of Fig.~\ref{fig4}(a).
Confirming our prediction, this shows that in this case, the coupling to the
ZGNR no longer favors a LS but leads to KS state with an enhanced
$T_K$. In Fig.~\ref{fig5}(b) we show $T^*$ as a function of the impurity position $h_n$, for the
two different chemical potential used in Fig.~\ref{fig5}(a). As we see, $T^*$
still decreases quite substantially for $\mu=-2.17t$ [squares (blue)] but for
$\mu=-2.28t$ [circle (red)] is almost as the impurity moves far away from the
edge. This can be understood based on the electronic states that contribute
more to the higher energy van Hove singularities are more extended across the ZGNR.

\begin{figure}[h]
\begin{center}
\includegraphics[scale=0.46]{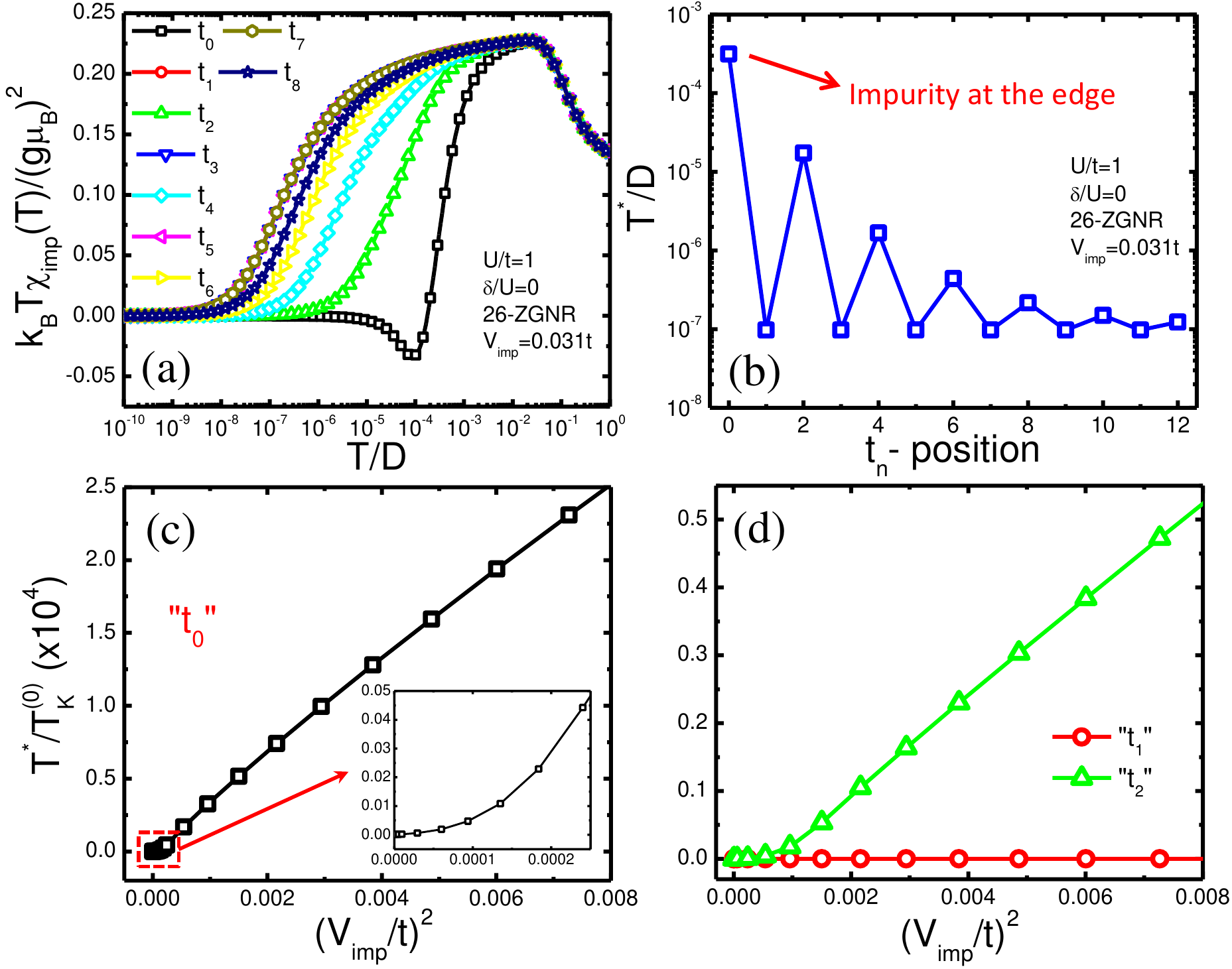}
\caption{(a) Magnetic moment vs $T$ for the impurity located at different
top site positions $t_{n}$. (b) Characteristic temperature, $T^*$, vs $t_n$ for
$V_\imp=0.031t$ [the same as in \ref{fig3}(d)]. (c) and (d) shows $T^*$ vs $(V_{\imp}/t)^2$ for
the impurity at the position $t_0$, $t_1$ and $t_2$, respectively.}
\label{fig6}
\end{center}
\end{figure}

\subsection{Top site adatom}
We now investigate if the two screening
regimes discussed above also occur for the adatom placed in a top site position.
The results obtained in this case are shown in Fig.~\ref{fig6}.
Fig.~\ref{fig6}(a) shows $k_BT\chi_\imp$ vs temperature for different
positions $t_{n}$ and for a fixed value of $V_\imp=0.031t$ [as in
Fig.~\ref{fig3}(c)]. Similar to the hollow site positions, we note that the
characteristic temperature $T^*$ (where $k_BT\chi_\imp$ drops to zero) increases
as the impurity adatom approaches to the edge of the ZGNR. Also, alike the hollow site case, a
dramatic change in shape of the $k_BT\chi_\imp$ curve occurs for $t_0$ as
compared to the others. This change of regime is accompanied by dramatic drop in
$T^*$ as shown in Fig.~\ref{fig6}(b), which shows $T^*$ vs $t_n$.
Interestingly, we note that $T^*$ exhibits a damping  oscillation as one move
the adatom away from the edge of the ZGNR. This behavior was not observed for
the hollow site positions. This difference can be understood as follows: in the
hollow site case, the adatom is connected to all six carbon atoms of the
hexagon, hence  the contribution from these neighboring carbons to the
corresponding hybridization function is somehow \textit{averaged}. Here, in
top site position, the total contribution to the hybridization function comes
solely from a single carbon atom. Therefore, a systematic oscillation with an
exponential decay of the LDOS across the ZGNR,\cite{Fujita} directly influences
the hybridization function $\Delta(E)$. In Fig.~\ref{fig6}(c)
we show $T^*$ vs $V_\imp^2$ for position $t_0$ and in \ref{fig6}(d) we
show the same for the positions $t_{1}$ and $t_2$. We clearly note a linear
behavior of $T^*$ for large $V_\imp$ for both $t_0$ and $t_2$. Also, for the
$t_0$ adatom position (and the subsequent $t_{2*n}$) we see an exponential
behavior of $T^*$ for small $V_\imp$. This is better appreciated in
the inset of the Fig.~\ref{fig6}(c) which shows a zoom of the region of small
$V_\imp$ for the $t_{0}$. For the $t_1$ position the $T^*$ is practically
insensitive to changes in $V_\imp^2$. These results show that the two LS and
KS screening regimes indeed occur for both hollow and top site positions,
whenever the adatom is placed close to the edge of the ZGNR.

\section{Conclusion}
We have studied the screening effect of a magnetic impurity  (adatom) on a ZGNR.
The system was described by an Anderson-like Hamiltonian where the adatom is
coupled to the ZGNR as well as to a metallic STM tip.
To access the low-temperature physics of the system we have
employed a numerical renormalization group approach that allows us to calculate
the relevant physical quantities. In particular, we have calculated the magnetic
moment of the system, through which we extract the characteristic temperature
$T^*$ below which the adatom magnetic moment is screened. We have analyzed
either the \emph{hollow} and \emph{top site} adatom configurations.  We
have found two screening regimes of the adatom magnetic moment: (1) a local
singlet (LS), when the adatom is \emph{strongly} coupled to the bound edge state
of the ZGNR and (2) a Kondo singlet (KS) in the \emph{weak} coupling case. The
system crosses over the LS to the usual KS either as the impurity is moved away
from the edge of the ZGNR or when its coupling $V_\imp$ to the ZGNR is small.
These two screening regimes are well defined by the behavior of the
characteristic temperature $T^*$ with $V_\imp$.
In the LS regime, $T^*$ increases linearly with $V_\imp^2$ whereas in the KS it
increases exponentially with $V_\imp^2$. We have shown that in the LS regime,
the linear dependence of $T^*$ with $V_\imp^2$ is consistent with a singlet
state formed between the magnetic moment of the impurity, and the one of an
electron in the bound edge state. Interestingly, the KS can be understood in
terms of an enhancement of the Kondo temperature as $V_\imp$ increases. In this
sense, in the LS regime the ZNGR state that is bound to the edge competes with
the Kondo screening of the adatom magnetic moment by the conduction electrons of
the STM tip, whereas in the KS case the ZGNR extended state cooperates with the
Kondo screening by the STM tip. Our results are important to the comprehension
of the Kondo physics in graphene nanoribbons and, given the relative simplicity
of the physical system studied here, we believe that these results can be
readily confirmed in STM experiments.

\section{Acknowledgments}
We thank R. \v{Z}itko for his assistance in the NRG
Ljubljana code.\cite{Ljubljana} We acknowledge financial support received from
CAPES, FAPEMIG, FAPERJ and CNPq.


\begin{thebibliography}{57}
\bibitem [{ada()}]{adatom}%
  \BibitemOpen
  \href@noop {} {}\bibinfo {note} {Here we have also calculated $\Delta(E=0)$
  for the adatom located at intermediate hollow positions $h_{1/2}$,
  $h_{3/2}$,...,$h_{7/2}$. Later, we will consider only at $h_{n}$, for $n$
  integer without loss of generality for the thermodynamical
  results.}\BibitemShut {Stop}%
\end{thebibliography}
%
\end{document}